\documentclass[11pt, letterpaper]{article} \usepackage[margin = 1in]{geometry}
\usepackage{amsmath, amsfonts, amssymb, float}
\usepackage{graphicx, subfigure, multirow}
\usepackage{capt-of}

\begin{document}
	
\title{Dissipative Particle Dynamics Simulation of Critical Pore Size in a Lipid Bilayer Membrane} %
\author{Clark Bowman\thanks{Division of Applied Mathematics, Brown University, Providence, Rhode Island 02912, USA}\ \thanks{\texttt{clark\_bowman@brown.edu}}
\and
Mark Chaplain\thanks{School of Mathematics and Statistics, University of St Andrews, St Andrews KY16 9SS, UK}\ \thanks{\texttt{majc@st-andrews.ac.uk}}
\and
Anastasios Matzavinos\footnotemark[1]\ \thanks{\texttt{matzavinos@brown.edu}}}

\maketitle

\begin{abstract} 
We investigate with computer simulations the critical radius of pores in a lipid bilayer membrane. Ilton et al. \cite{PhysRevLett.117.257801} recently showed that nucleated pores in a homopolymer film can increase or decrease in size, depending on whether they are larger or smaller than a critical size which scales linearly with film thickness. Using dissipative particle dynamics, a particle-based simulation method, we investigate the same scenario for a lipid bilayer membrane whose structure is determined by lipid-water interactions. We simulate a perforated membrane in which holes larger than a critical radius grow, while holes smaller than the critical radius close, as in the experiment of \cite{PhysRevLett.117.257801}. By altering system parameters such as the number of particles per lipid and the periodicity, we also describe scenarios in which pores of any initial size can seal or even remain stable, showing a fundamental difference in the behavior of lipid membranes from polymer films.
\end{abstract} 

\section{Introduction\label{sec_intro}}

A recent article by Ilton et al. \cite{PhysRevLett.117.257801} examined the evolution of pores in a model membrane constructed from polystyrene. A tightly focused laser was used to create a temperature gradient in the film, decreasing the local surface tension and driving the formation of a hole whose size was determined by the power and exposure time of the laser. It was found that holes below a critical size $r_c$ sealed, while larger holes began to increase in size.

The polymer films of \cite{PhysRevLett.117.257801} are easily observable via traditional microscopy techniques due to their large size (the thinnest homopolymer film studied was $\sim$100 nm thick, with most films on the order of 1 $\mu m$). In contrast, a typical lipid membrane, as might be found in a biological cell, is much smaller ($\sim$10 nm thick). Pore formation and evolution in a lipid membrane thus has a number of experimental difficulties owing to the small spatial and temporal scales. In this work, we instead use a simulated lipid membrane to study the dynamics in question. The simulated system is computationally feasible and reproduces many of the results observed in \cite{PhysRevLett.117.257801}.

Simulating dynamics on the scale of individual atoms and molecules (such as lipids) is often done with molecular dynamics, a class of commonly-used simulation methods which operate by numerically solving Newton's equations of motion for each particle. Because particles are simulated individually, the computational cost of such simulations becomes infeasible at even moderate scales. This is particularly evident when simulating particles or structures immersed in fluids (e.g., membranes), whose characteristic time of motion often differs significantly from that of the solvent.

To explicitly simulate a lipid bilayer membrane, we employ dissipative particle dynamics (DPD), a modification of molecular dynamics which reduces computational complexity by aggregating small groups of like atoms or molecules into single ``dissipative particles.'' Extensive comparisons with molecular dynamics and Navier-Stokes simulations have shown that, with the proper choice of intermolecular forces, DPD simulations maintain the correct hydrodynamic behavior in a wide range of spatial and temporal scales \cite{Espanol1995b,doi:10.1063/1.2018635,shardlow:2006}. The result is a method which explicitly models the solvent (via coarse-grained dissipative particles) while being computationally feasible at the scale of large biological structures \cite{moeendarbary2010}.

We begin by describing the explicit formulation of DPD in Section \ref{sec_dpd}. The construction of the membrane and measurement techniques are detailed in Section \ref{sec_lipid}. Sections \ref{sec_periodic} and \ref{sec_finite} present the simulation results. Finally, the results are contextualized in Section \ref{sec_discussion}.

\section{The DPD Simulation Method\label{sec_dpd}}

Let the mass, velocity, and position of dissipative particle $i$ be given by $m_i$, $\textbf{v}_i$, and $\textbf{r}_i$, respectively. The DPD equation of motion for particle $i$ comprises three pairwise contributions:
$$m_i\frac{d\textbf{v}_i}{dt} = \sum\limits_{j\neq i}\textbf{F}_{ij}^C+\sum\limits_{j\neq i}\textbf{F}_{ij}^D+\sum\limits_{j\neq i}\textbf{F}_{ij}^R.$$

$\textbf{F}_{ij}^C$ is a conservative force deriving from a potential exerted on particle $i$ by particle $j$, similar to the usual pairwise forces implemented in molecular dynamics schemes. Here, we adopt the common explicit form
$$\textbf{F}_{ij}^C = \begin{cases} a_{ij}\left(1-\frac{r_{ij}}{r_c}\right)\hat{\textbf{r}}_{ij} &\mbox{if } r_{ij} < r_c \\
0 & \mbox{else,}\end{cases}$$
in terms of a conservative coefficient $a_{ij}$, the inter-particle displacement $\textbf{r}_{ij} = \textbf{r}_i - \textbf{r}_j$ (with magnitude $r_{ij}$ and unit vector $\hat{\textbf{r}}_{ij}$), and a cutoff radius $r_c$. The DPD conservative force is a soft potential (i.e., does not diverge as $r_{ij} \to 0$), and so particles can overlap or even occupy the same point in space, corresponding to the notion of DPD particles as coarse-grained clusters of smaller atoms or molecules.

The dissipative force $\textbf{F}_{ij}^D$ and random force $\textbf{F}_{ij}^R$ function as a thermostat and are given by
\begin{eqnarray*}
\textbf{F}_{ij}^D &=& -\gamma_{ij}\omega^D(r_{ij})\left(\hat{\textbf{r}}_{ij}\cdot \textbf{v}_{ij}\right)\hat{\textbf{r}}_{ij}\\
\textbf{F}_{ij}^R &=& -\sigma_{ij}\omega^R(r_{ij})\theta_{ij}\hat{\textbf{r}}_{ij}
\end{eqnarray*}
where $\omega^D(\cdot)$ and $\omega^R(\cdot)$ are position-dependent weight functions, $\textbf{v}_{ij} = \textbf{v}_i - \textbf{v}_j$, and $\gamma_{ij}$ and $\sigma_{ij}$ are the dissipative and random strengths, respectively. Noise is introduced by the Gaussian white-noise term $\theta_{ij}$, which satisfies the stochastic conditions
$$\langle\theta_{ij}(t)\rangle = 0\hspace{0.3cm}\text{and}\hspace{0.3cm}\langle\theta_{ij}(t)\theta_{kl}(\tau)\rangle = (\delta_{ik}\delta_{jl} + \delta_{il}\delta_{jk})\delta(t-\tau).$$

In order to ensure conservation of momentum, it is assumed that the noise terms are symmetric in $i$ and $j$, i.e., $\theta_{ij} = \theta_{ji}$. Espa\~{n}ol and Warren \cite{Espanol1995b} provided an additional condition in order to preserve the invariant distribution of the system with conservative forces alone, namely, the fluctuation-dissipation relation
$$\omega^D(r_{ij}) = \left[\omega^R(r_{ij})\right]^2\hspace{0.3cm}\text{and}\hspace{0.3cm}\sigma_{ij}^2 = 2\gamma_{ij} k_BT,$$
where $k_B$ is the Boltzmann constant and $T$ the equilibrium temperature. In net, the DPD equations of motion can then be written as the following set of coupled stochastic differential equations:

\begin{eqnarray}
\label{dpdeqnsofmotion}
\begin{cases} d\textbf{r}_i = \textbf{v}_idt\\
d\textbf{v}_i = \frac{dt}{m_i}\sum\limits_{j\neq i}\left(\textbf{F}_{ij}^C(\textbf{r}_{ij})-\gamma_{ij}\left[\omega^R(r_{ij})\right]^2\left(\hat{\textbf{r}}_{ij}\cdot \textbf{v}_{ij}\right)\hat{\textbf{r}}_{ij}\right) + \frac{1}{m_i}\sum\limits_{j\neq i}\left(\sqrt{2\gamma_{ij} k_BT}\right)\omega^R(r_{ij})\hat{\textbf{r}}_{ij}dW_{ij},\end{cases}
\end{eqnarray}
where $W_{ij} = W_{ji}$ is a Wiener process for each $i, j$. To proceed with the DPD simulation, the system is then stepped forward with a numerical solver. Here, we use the DPD velocity-Verlet scheme: given positions $\textbf{r}^n$ and velocities $\textbf{v}^n$ at step $n$ and a timestep $\Delta t$, compute the half-step velocities

$$\textbf{v}_i^{n+1/2} = \textbf{v}_i^n + \frac{1}{2m_i}\Big(\textbf{F}_i^C(\textbf{r}^n)\Delta t
+\textbf{F}_i^D(\textbf{r}^n,\textbf{v}^n)\Delta t+\textbf{F}_i^R(\textbf{r}^n)\sqrt{\Delta t}\Big),$$
then calculate the next step as
\begin{eqnarray*}\textbf{r}_i^{n+1}&=&\textbf{r}_i^n+\textbf{v}_i^{n+1/2}\Delta t,\\
\textbf{v}_i^{n+1}&=&\textbf{v}_i^{n+1/2}+\frac{1}{2m_i}\Big(\textbf{F}_i^C(\textbf{r}^{n+1})\Delta t
+\textbf{F}_i^D(\textbf{r}^{n+1},\textbf{v}^{n+1/2})\Delta t+\textbf{F}_i^R(\textbf{r}^{n+1})\sqrt{\Delta t}\Big).
\end{eqnarray*}

This scheme is very similar to traditional velocity-Verlet, with the exception that the dissipative force term $\textbf{F}^D$ must be calculated twice per step since it is both position- and velocity-dependent.

\section{Lipid Bilayer Simulation\label{sec_lipid}}

It remains to specify the coefficients of Eq. (\ref{dpdeqnsofmotion}). Because there is a choice of scale for the coarse-graining, DPD coefficients are usually specified for the nondimensionalized system. In their foundational paper on DPD, Groot and Warren \cite{groot} found that the dimensionless compressibility of water could be matched by using the conservative coefficient $a_{ij} = 25.0$, dissipative coefficient $\gamma_{ij} = 4.5$, cutoff radius $r_c = 1.0$, and numerical density $\rho = 3$ for unit-mass DPD particles. To model a lipid bilayer membrane, we additionally introduce particles with modified coefficients to model the lipids, here referred to as ``Head'' and ``Tail'' particles. The pairwise $a_{ij}$ and $\gamma_{ij}$, chosen to be similar to existing literature on DPD membranes \cite{doi:10.1063/1.2424698, 10.1016/j.bpj.2008.11.073, doi:10.1063/1.1498463} and to reproduce mesoscopic properties such as lateral fluidity and a stable bilayer structure \cite{B406433J}, are shown in Table \ref{tab_coeff}.

\begin{table}[!ht]
\label{tab_coeff}
    \caption{Pairwise Coefficients for DPD Membrane}
		\ \\
    \begin{minipage}{.5\linewidth}
      \centering
        \begin{tabular}{|c|ccc|}
\hline
$a_{ij}$ & Head & Tail & Water\\
\hline
Head & 25.0 & 50.0 & 35.0\\
Tail & 50.0 & 15.0 & 75.0\\
Water & 35.0 & 75.0 & 25.0\\
\hline
\end{tabular}
    \end{minipage}%
    \begin{minipage}{.5\linewidth}
      \centering
        \begin{tabular}{|c|ccc|}
\hline
$\gamma_{ij}$ & Head & Tail & Water\\
\hline
Head & 4.5 & 9.0 & 4.5\\
Tail & 9.0 & 4.5 & 20.0\\
Water & 4.5 & 20.0 & 4.5\\
\hline
\end{tabular}
    \end{minipage} 
\end{table}

\begin{figure}[ht]
\centering
\textbf{Lipid Schematic \& Bilayer Initialization}
\begin{tabular}{cc}
\subfigure[]{\includegraphics[width=.22\linewidth]{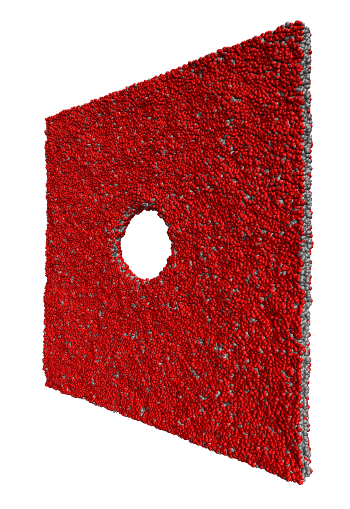}} & \subfigure[]{\includegraphics[width=.4\linewidth]{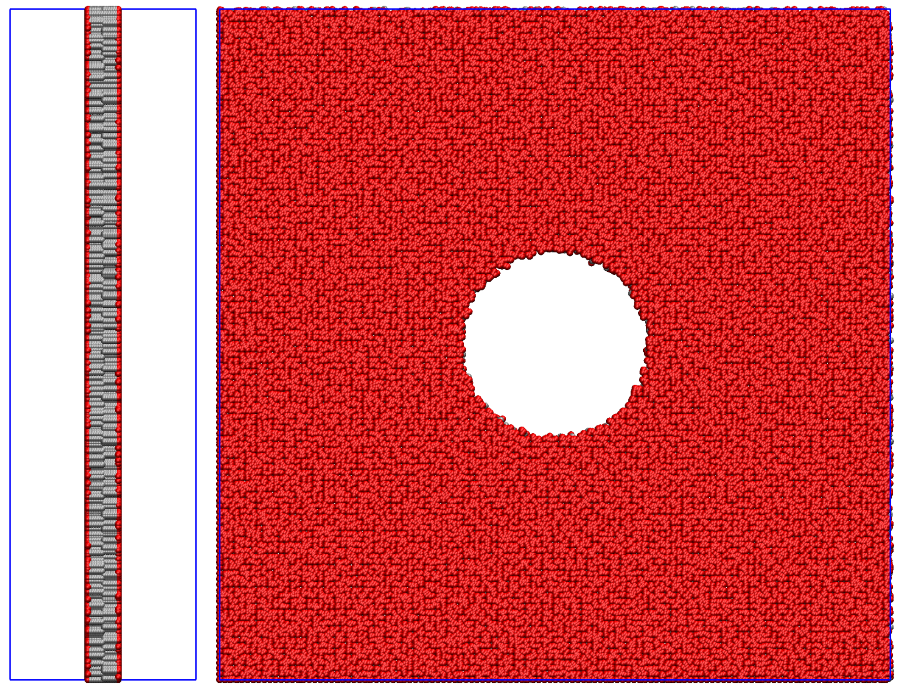}}\\
\subfigure[]{\includegraphics[width=.275\linewidth]{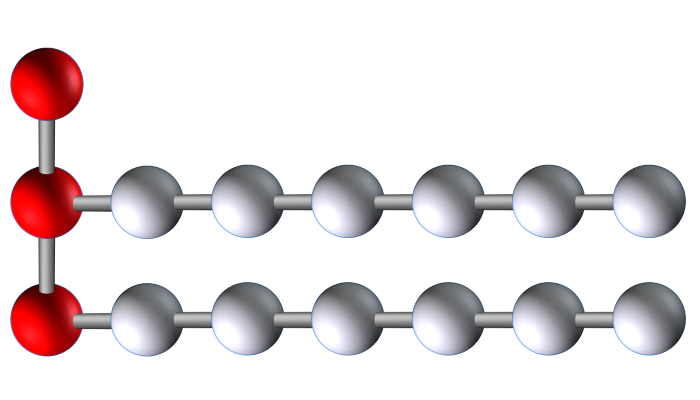}} & \subfigure[]{\includegraphics[width=.2\linewidth]{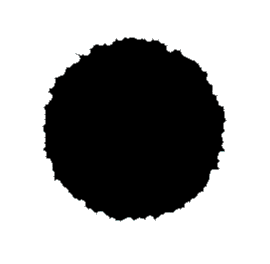}}\\
\end{tabular}
\caption{\label{fig_setup}(a) 3D view of simulated membrane with pore. Fluid particles (not shown) fill the space on both sides of the membrane and in the pore. (b) Side and front view of initial membrane configuration. Blue lines at boundary show the simulation box. (c) Lipid with three heads and two tails comprising six tail particles each. Bonds are shown as thin rods. Three-body potentials are imposed among triplets with consecutive horizontal bonds. (d) Binary matrix (value 1 shown in black) resulting from smoothing and thresholding the front view in (b).}
\end{figure}

Head and tail particles are connected as in Figure \ref{fig_setup}(c) by harmonic bonds with dimensionless energy $E_{ij} = 64(r_{ij} - 0.5)^2$, so that the resting length of a bond is 0.5. Along the tails, three-body potentials $E_{ijk} = 20(1+\cos{\phi_{ijk}})$ are introduced to provide stiffness, where $\phi_{ijk}$ is the angle between bond $ij$ and bond $jk$. Copies of the lipids are then placed in two layers at the vertices of a square lattice, with the tails facing inward. The outside of the membrane is initialized with water particles placed on a cubic lattice with numerical density $\rho = 3$. Periodic boundary conditions are imposed around the simulation box.

For the first simulation, the periodic simulation box of $144 \times 144 \times 40$ was filled entirely in the first two dimensions with a bilayer membrane comprising 28,700 lipids with three head particles and two tails of six particles each, so that the side length of the square lattice on which lipids were initialized was approx. 1.202. The number of lipids was chosen experimentally to achieve stability in the membrane. To simulate a pore, all lipids intersecting an orthogonal cylinder of fixed radius were removed, and the cylinder was included in the initialization region for fluid particles. The resulting setup at time zero can be seen in Figure \ref{fig_setup}(b).

In each simulation, the time evolution of the pore was observed using DPD velocity-Verlet with a temperature $T = 1.0$ and timestep $\Delta t = 0.005$. To track the size of the pore, the locations of all lipids were recorded every 1000 steps. Images of the system were created by rendering spheres of radius 0.5 at the location of every head and tail particle, then projecting the 3D system into a 2D plane tangent to the membrane surface. The resulting images were then smoothed and thresholded, resulting in binary matrices $M_i$ with value 1 within the pore and 0 without, as in Figure \ref{fig_setup}(d). Finally, the 2D center of mass position $\bar{\textbf{x}}_i$ was computed, and the sum $\sum\limits_{\textbf{x}}M_i(\textbf{x})\left|\textbf{x}-\bar{\textbf{x}}_i\right|$ was calculated. For a circle of radius $R$, the result should be approximated by the integral

$$\int\limits_0^{2\pi}\int\limits_0^R|r|rdrd\theta = \frac{2\pi}{3}R^3,$$

and so the approximate radius of the pore is given by
\begin{equation}R_i \approx \left(\frac{3}{2\pi}\sum\limits_{\textbf{x}}M_i(\textbf{x})\left|\textbf{x}-\bar{\textbf{x}}_i\right|\right)^{1/3}.\label{l_rad}\end{equation}

To compute the membrane width, the 3D representation was instead projected into a plane bisecting the membrane (i.e., an orthographic side view) as in the side view of Figure \ref{fig_setup}(b). For each row in the projected image, the number of non-background pixels was summed, approximating the width of the membrane at that location; to reduce the effect of the natural thermal fluctuations in the membrane surface, the minimum of all row widths was used as the membrane width for that image. The resulting value was time-averaged over 20 images (one every 1000 steps) to obtain a reference value $h$ for the membrane width.

\section{Pores in Periodic Membranes\label{sec_periodic}}
\subsection{Existence of a Critical Radius\label{sec_periodic1}}

\begin{figure}[ht]
\centering
\textbf{Time Evolution of Membrane Pores (II)}\\
\includegraphics[width=\linewidth]{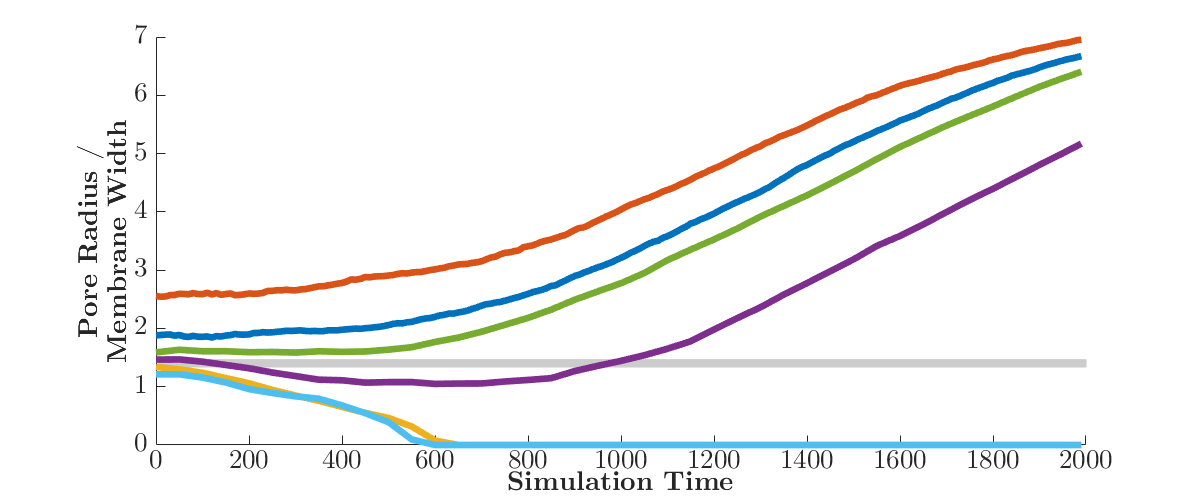}
\captionof{figure}{\label{fig_6lipid_b}Membrane pore size over time as calculated by Eq. (\ref{l_rad}). Size ($y$ axis) is shown in units of the average membrane width, $h = 7.98$. Several initial sizes are shown, ranging from $r_0/h = 1.21$ (bottom) to $r_0/h = 2.56$ (top). The shaded region denotes possible values of the critical ratio $r_c/h$ below which pores seal -- $1.34 < r_c/h < 1.47$.}
\end{figure}

Results for pores of various initial radii can be seen in Figure \ref{fig_6lipid_a}. The behavior of the pore is a function of its radius: sufficiently small pores begin to shrink and eventually seal entirely, while sufficiently large pores begin expanding, eventually severing the membrane. The time-evolving radius of each pore, as measured by Eq. (\ref{l_rad}), is shown in Figure \ref{fig_6lipid_b}.

The growth of supercritical pores proceeds in roughly exponential fashion, in agreement with the findings for homopolymer films in \cite{PhysRevLett.117.257801}. For very large times ($t > 1000$), the pore approaches the order of the simulation box, and so begins interacting with itself across the periodic boundary, resulting in slower growth. The subcritical pore, which closed around $t = 400$, yielded a stable membrane for the remainder of the simulation. The observed critical ratio $1.34 < r_c/h < 1.47$ is larger than the ratio $r_c/h = \pi/4$ derived by Ilton et al. for a homopolymer film., suggesting an additional free energy cost for pore formation due to the lipid structure. Note that we define the ratio in terms of the initial size $r_0$, meaning it is possible a pore which evolves to be smaller than the critical radius may still expand as $t \to \infty$.

\begin{tabular}{rcccc}
\multicolumn{5}{c}{\textbf{Time Evolution of Membrane Pores (I)}} \\
 & $t = 0$ & $t = 200$ & $t = 400$ & $t = 1000$ \\
\raisebox{1.2cm}{$r_0 = 10$} & \includegraphics[width=.18\linewidth]{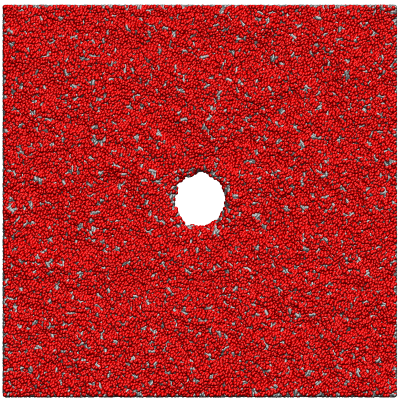} & \includegraphics[width=.18\linewidth]{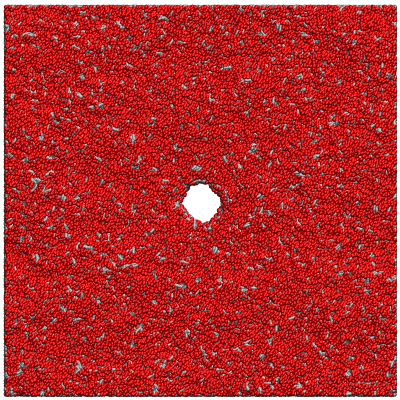} & \includegraphics[width=.18\linewidth]{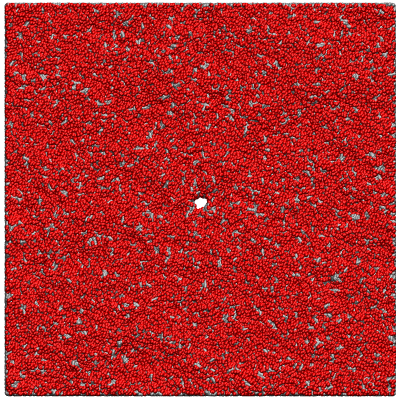} & \includegraphics[width=.18\linewidth]{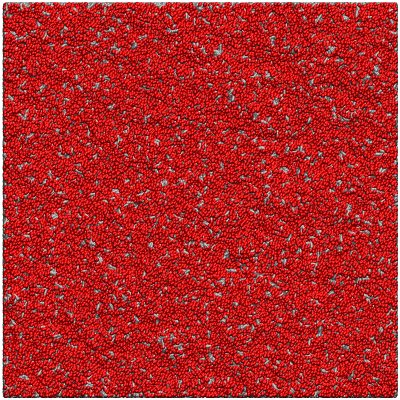}\\
\raisebox{1.2cm}{$r_0 = 15$} & \includegraphics[width=.18\linewidth]{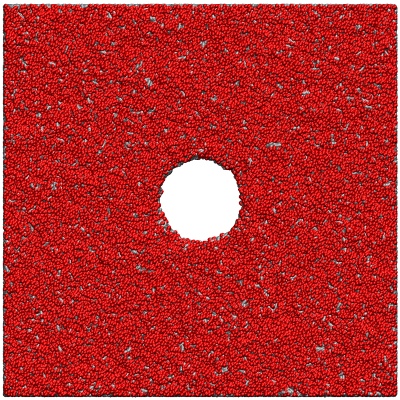} & \includegraphics[width=.18\linewidth]{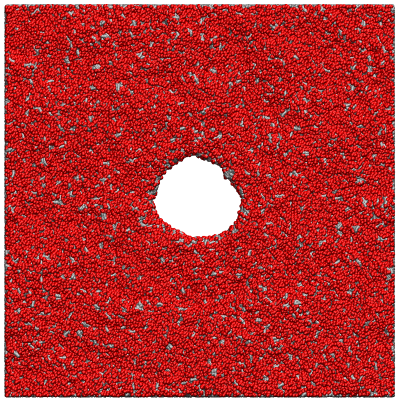} & \includegraphics[width=.18\linewidth]{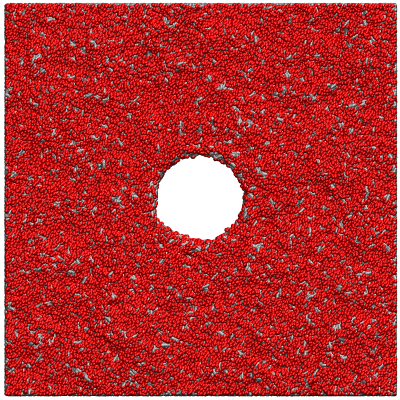} & \includegraphics[width=.18\linewidth]{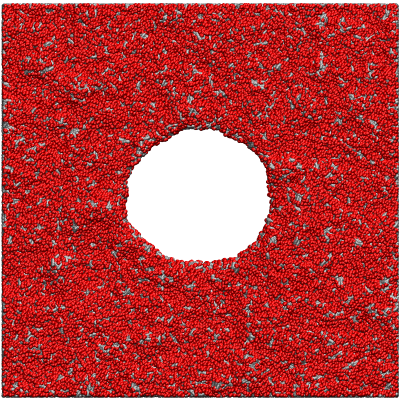}\\
\raisebox{1.2cm}{$r_0 = 20$} & \includegraphics[width=.18\linewidth]{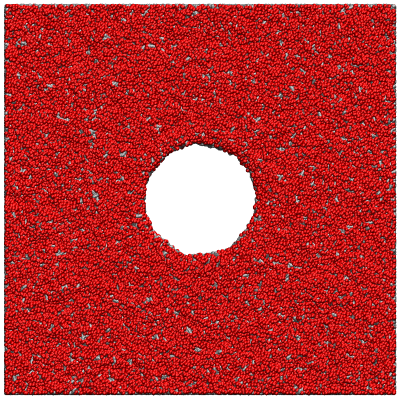} & \includegraphics[width=.18\linewidth]{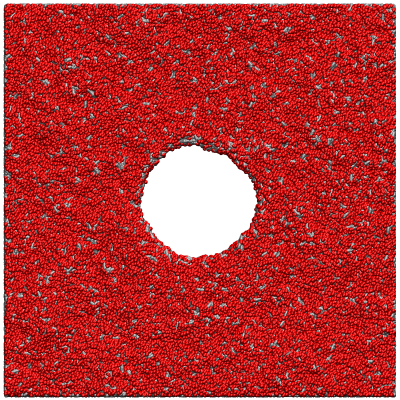} & \includegraphics[width=.18\linewidth]{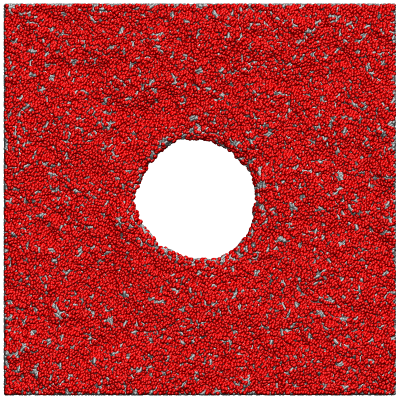} & \includegraphics[width=.18\linewidth]{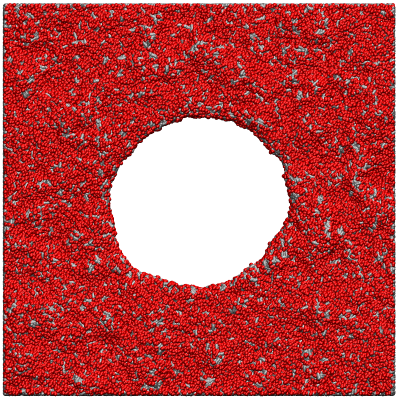}\\
\end{tabular}
\captionof{figure}{\label{fig_6lipid_a}Rendered images of pores of three different initial sizes (rows). Four images throughout the simulation are shown, progressing from left to right. Pores shrink or grow exponentially depending on the initial pore radius.}

\newpage

\begin{figure}[ht]
\centering
\textbf{High-Density Membrane Pore Evolution}\\
\includegraphics[width=\linewidth]{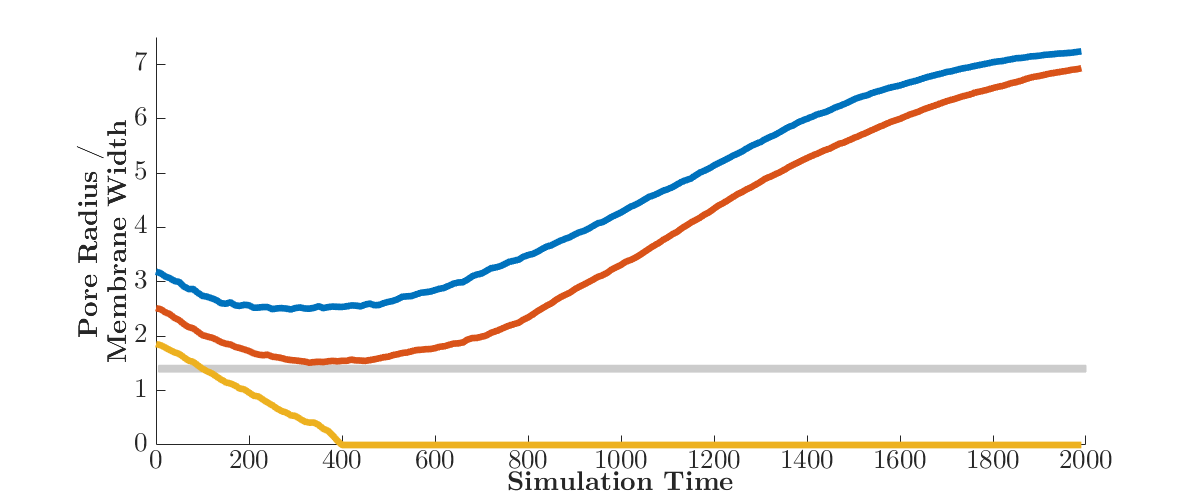}
\captionof{figure}{\label{fig_6lipid_hd}Pore size over time via Eq. (\ref{l_rad}) for a membrane with increased lipid density. Three initial sizes are shown: top: $r_0/h = 3.19$, middle: $r_0/h = 2.52$, and bottom: $r_0/h = 1.86$. The shaded region corresponds to the possible range of $r_c/h$ derived in Figure \ref{fig_6lipid_b}.}
\end{figure}

A second set of simulations examined the effect of increasing internal pressure in the membrane. The number of lipids in the periodic simulation box was increased by 4.4\% to 29,970, resulting in a stable membrane with a lipid excess. For short times, a pore opened in the modified membrane begins to close regardless of pore size as the membrane relieves internal pressure. As seen in Figure \ref{fig_6lipid_hd}, the existence of a critical ratio remains in this scenario, as sufficiently large pores reverse the initial collapse and expand exponentially as in the first simulation. This behavior is governed by equilibrium size, rather than initial size; the pore with $r_0/h = 1.88$, well above the critical ratio, closes regardless in this new scenario despite having expanded in the initial case of Figure \ref{fig_6lipid_b}.

\subsection{Membranes of Varied Thickness\label{sec_periodic2}}

The curvature of the membrane around the lip of the pore decreases with the membrane width $h$; increasing the thickness should decrease the energy cost of pore formation, thereby increasing the critical radius $r_c$. It was shown in \cite{PhysRevLett.117.257801} that for a homogeneous film, where the free energy cost of pore formation is derived entirely from edge tension, the scaling should be linear as $r_c = h\pi/4$. To examine this scaling for the simulated lipid membrane, we changed the membrane thickness by altering the number of particles per lipid tail.

Initially, the tail length was increased from 6 to 7 particles. The resulting membrane was found to be stable and at equilibrium with 29,200 lipids in the same simulation box, i.e., a lipid density increase of 1.74\%. The resulting membrane had a width of approximately $h = 8.63$ (8.15\% thicker than the original membrane).

\begin{figure}[ht]
\centering
\textbf{Membrane Pore Evolution, 7 Lipids per Tail}\\
\includegraphics[width=0.9\linewidth]{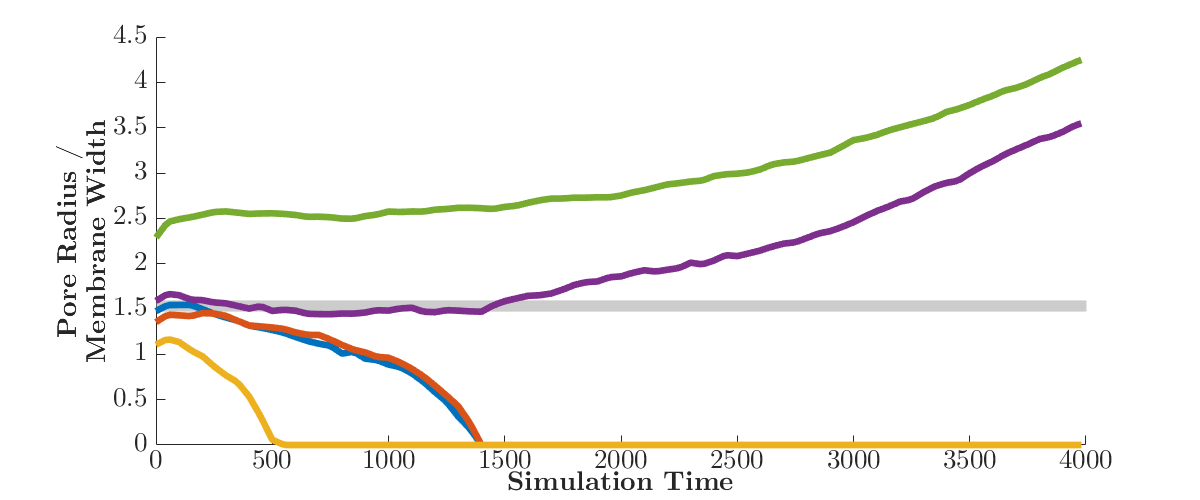}
\captionof{figure}{\label{fig_7lipid}Membrane pore size over time from Eq. (\ref{l_rad}). Size ($y$ axis) uses the new average membrane width $h = 8.63$. The shaded region corresponds to the new range for the critical radius, $1.48 < r_c/h < 1.59$.}
\end{figure}

Results for this set are shown in Figure \ref{fig_7lipid}. The single additional tail particle induced an upward shift in the critical ratio, to $1.48 < r_c/h < 1.59$, an increase of between 0.68\% -- 18.66\% from the 6 particle case. In addition, there was a notable change in the timescale on which pores evolved away from the critical region. In particular, the pore initialized at $r_0/h = 1.59$ was nearly stable for the entire duration of the previous simulations $(t \approx 2000)$ before eventually opening up into supercritical growth. Figure \ref{fig_7lipid} also demonstrates the stochastic nature of the dynamics at this spatial scale: in these realizations, the pore with initial size $r_0/h = 1.48$ closed faster than an initially smaller pore with radius $r_0/h = 1.36$.

\begin{figure}[ht]
\centering
\textbf{Membrane Pore Evolution, 12 Lipids per Tail}\\
\includegraphics[width=0.9\linewidth]{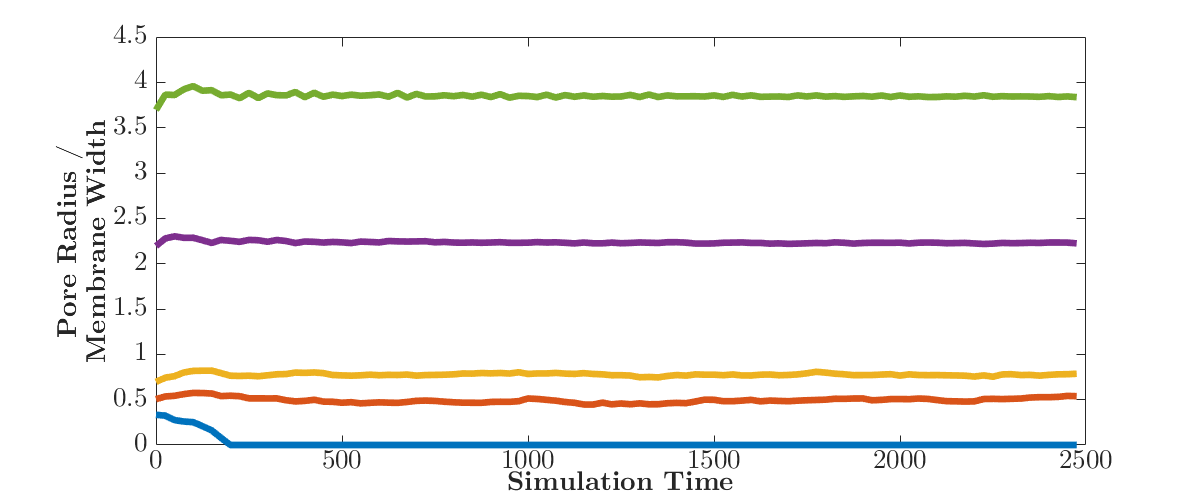}
\captionof{figure}{\label{fig_12lipid}Pore size over time via Eq. (\ref{l_rad}) for a membrane with 12 particles per lipid tail. Only the smallest pore simulated was not stable for the duration of the simulation. This membrane had a width of $h = 13.49$, or $69.05\%$ thicker than the membrane of Figures \ref{fig_6lipid_b}-\ref{fig_7lipid}.}
\end{figure}

Next, the tail length was increased further to 12 particles. The resulting time series can be seen in Figure \ref{fig_12lipid}. For very small pores, the behavior was unchanged, with the hole quickly being sealed. For pores of moderate size, the doubling of lipid tail length afforded significantly increased stability -- above some size threshold, all pores in the membrane are stable indefinitely, showing a marked contrast with the behavior of the membranes of Figures \ref{fig_6lipid_b}--\ref{fig_7lipid}. The largest simulated pore ($r_0 = 50$) required the simulation box be expanded to $240 \times 240 \times 40$ to avoid interference from periodic boundary effects.

Tail lengths of 8 and 9 lipids were also considered, but were found to produce nearly identical results to the case of 12 lipids per tail. The simulated bilayer membrane thus exhibits a sharp change in stability as the number of lipids per tail increases from 7 to 8. 

\section{Pores in Finite Membranes\label{sec_finite}}

To further understand this change in stability, we finally considered the case of a finite membrane, i.e., a free-floating square patch of membrane of finite size. To simulate such a membrane, lipids placed on a lattice in the initialization phase were truncated a fixed distance of 5 units away from the edge of the simulation box. The empty space left by truncating the membrane was included in the region of initialization for fluid particles.

\begin{figure}[ht]
\centering
\textbf{Time Evolution of Perforated Finite Membranes}
\vspace{0.4cm}

\begin{tabular}{rccccrccc}
\multirow{2}{*}{\raisebox{-1.5cm}{(a)}} & \includegraphics[width=0.11\linewidth]{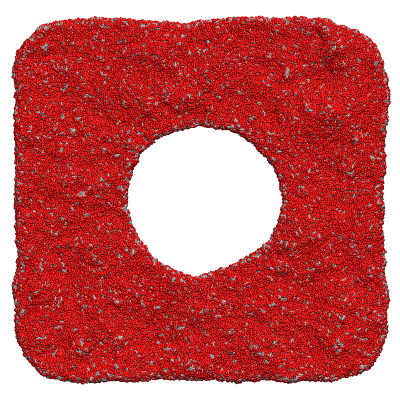} & \includegraphics[width=0.11\linewidth]{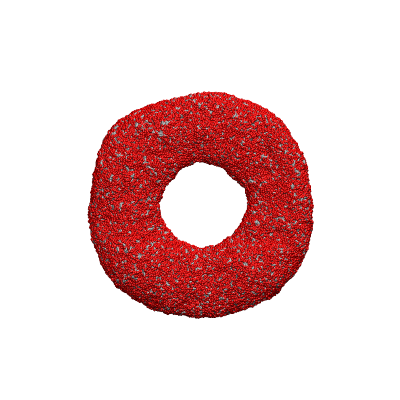} &\includegraphics[width=0.11\linewidth]{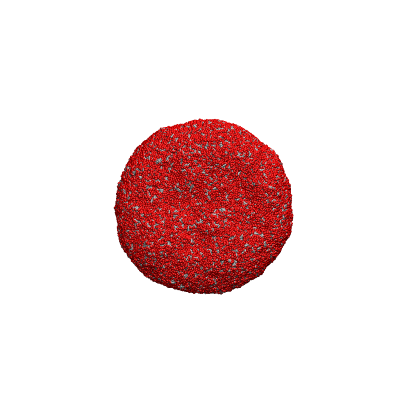} & & \multirow{2}{*}{\raisebox{-1.5cm}{(b)}} & \includegraphics[width=0.11\linewidth]{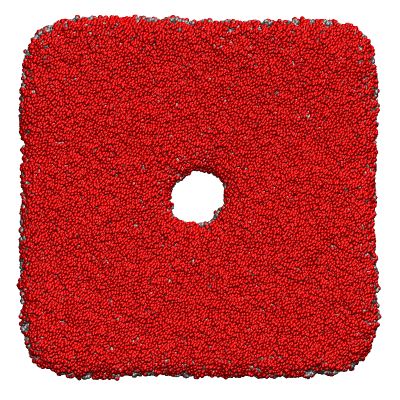} & \includegraphics[width=0.11\linewidth]{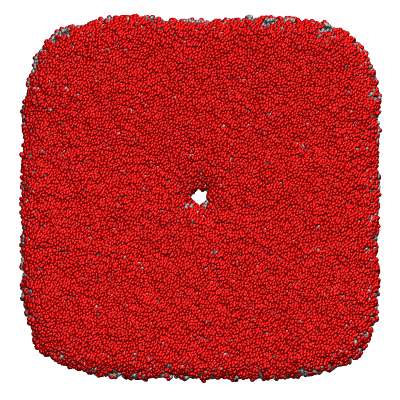} &\includegraphics[width=0.11\linewidth]{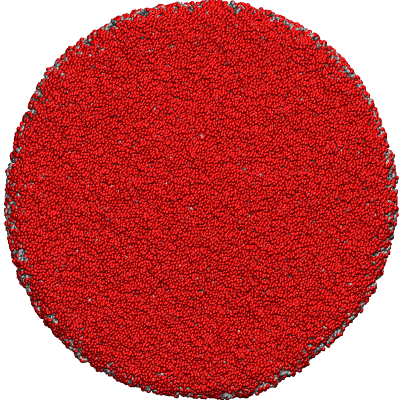}\\
 & \includegraphics[width=0.11\linewidth]{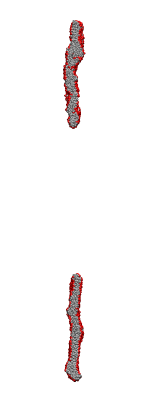} & \includegraphics[width=0.11\linewidth]{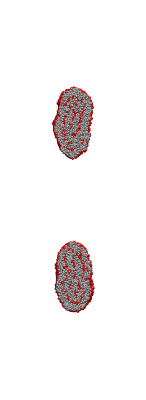} & \includegraphics[width=0.11\linewidth]{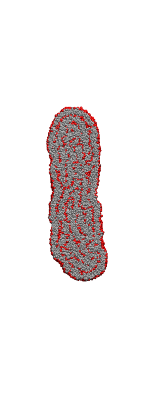} & \hspace{0.2cm} & & \includegraphics[width=0.11\linewidth]{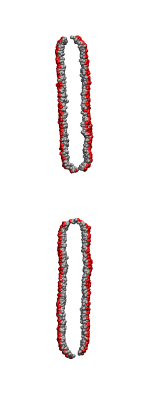} & \includegraphics[width=0.11\linewidth]{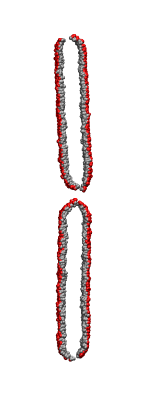} & \includegraphics[width=0.11\linewidth]{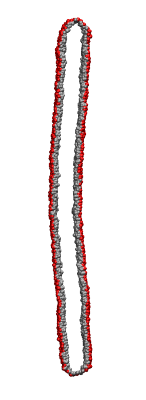}\\
\end{tabular}
\caption{\label{fig_finite}Rendered images of simulation results for truncated membranes; time increases left to right. Front and internal side views for the original membrane with 6 particles per lipid tail are shown in (a), while the modified membrane with 12 particles per lipid tail is shown in (b). The original membrane seals its pore but simultaneously transitions into a layered cluster. Conversely, the thicker membrane seals its pore but remains in a stable finite disc indefinitely. The interior of the membrane in (b) comprises only tail particles and is not shown.}
\end{figure}

All simulated finite membrane pores invariably sealed, regardless of initial radius. The membrane with 6 particles per lipid tail, which formerly exhibited a critical pore radius, transitioned through a metastable torus configuration to a layered cluster (see Figure \ref{fig_finite}). The membrane with 12 particles per lipid tail, whose pores were stable above a small threshold radius, sealed its pore but remained stable in a finite bilayer disc. We hypothesize that the existence of a critical radius of pores in the periodic case corresponds directly to the stability of the finite membrane; a stable finite membrane prevents pores from expanding regardless of their initial size.

\vspace{1cm}

\section{Discussion\label{sec_discussion}}

Many of the experimental findings about the polymer films of \cite{PhysRevLett.117.257801} were also observed in our dissipative particle dynamics simulation of a bilayer membrane. In particular, the existence of a critical pore radius was observed in the periodic simulations when membrane tails comprised 6--7 monomers. Ilton et al. explain the mechanism for such a phenomenon by writing the energy cost of pore formation $\Delta G(r)$ as a function of the pore radius: $(A-2\pi r^2)\gamma$, in terms of the surface area $A$ of the pore edge and the per-area surface tension of the film $\gamma$. Modeling the pore edge as the inner half-surface of a regular torus with diameter $h$ (the membrane width), the edge surface area $A$ is given by $\pi^2hr-\pi h^2.$ The resulting cost $\Delta G(r)$ is a concave function maximized when $\frac{\partial}{\partial r}\Delta G(r) = (\pi^2h - 4\pi r)\gamma = 0,$ yielding a critical radius $r_c = h\pi/4$ which scales linearly with the thickness $h$ of the film. The cost $\Delta G(r)$ is a barrier for pore formation: once the critical size is reached, further expansion of the pore begins to reduce the free energy.

This expression ignores any energy cost associated with the molecular structure of the membrane; the authors also describe a modified argument for a diblock film, which has been used as a simple model of a lipid bilayer membrane in theoretical work \cite{PhysRevE.88.012718}. Due to the additional cost of rearranging molecules on the curved surface around the pore, they predict a critical radius larger than for a homopolymer film by a factor proportional to the nondimensional curvature $L/h$, where $L$ is the equilibrium thickness of the lamellar layers. Our simulation results agree in this respect: critical radii of the lipid membranes with 6--7 monomer tails were found to be $1.34h < r_c < 1.47h$ and $1.48h < r_c < 1.59h$, respectively, compared to the homopolymer $r_c = h\pi/4 \approx 0.79h$.

There also exist significant differences between our simulations and the work of Ilton et al., most notably in the case of the thicker membrane. As the diblock copolymer correction term scales with the nondimensional curvature, its influence should decay as the membrane width increases, reaching the same limit of $r_c/h = \pi/4$ as $h \to \infty$. In contrast, our simulations of a thicker lipid bilayer membrane showed pores above a certain size to be stable indefinitely, i.e., a critical radius above which pores expanded no longer existed. This suggests that the structure of a lipid bilayer membrane is fundamentally different than the structure of a diblock copolymer. Unlike the polystyrene film and PS-b-PMMA diblock copolymer of \cite{PhysRevLett.117.257801}, whose critical radius was a continuous function of membrane width, the simulated bilayer membrane exhibits a phase transition from a regime where critical radius relates to thickness ($\leq 7$ monomer lipid tails) to a regime where arbitrarily large pores are stable ($\geq 8$ monomer lipid tails).

Our simulations additionally provide insight into the time scale on which growth occurs. Although pores above the critical radius (in simulations where a critical radius existed) were found to increase in size roughly exponentially, the time scale of the exponential growth was markedly different between the membranes with tail lengths of 6 and 7 particles. To compare these timescales, the time series for the radius of the smallest supercritical pore was fit to an exponential $r_0e^{t/\tau}$ in terms of the characteristic time $\tau$. The thinner membrane (\ref{sec_periodic1}) was found to have a characteristic time of $\tau_1 \approx 900$, while the slightly thicker membrane (\ref{sec_periodic2}) was found to have a characteristic time of $\tau_2 \approx 2000$. The time scale of pore evolution for lipid bilayer membranes is thus significantly affected by the structure/width, potentially in addition to chemical properties (in this context, the force coefficients for particle interaction, which were not changed between simulations).

Since the DPD method explicitly models the solvent and does not make equilibrium assumptions, it is also suitable for examining the behavior of lipid membrane pores in the presence of fluid flows or pressure gradients, such as those observed in biological cells during movement. Future work can examine the stability and dynamics of such pores in a variety of contexts of interest in cell biology.

\section*{Acknowledgments}

Part of this research was conducted using computational resources and services at the Center for Computation and Visualization, Brown University. CB was partially supported by the NSF through grant DMS-1148284. AM was partially supported by the NSF through grants DMS-1521266 and DMS-1552903. 

\bibliographystyle{plain}
\bibliography{main.bib}

\end{document}